\documentclass[12pt]{iopart}
\usepackage{graphicx}
\begin{document}

\title[Entropy of the Nordic electricity market]{Entropy of the Nordic electricity market: anomalous scaling, spikes, and mean-reversion}

\author{J Perell\'o$^1$, M Montero$^1$, L Palatella$^2$, I Simonsen$^3$ and J Masoliver}

\address{$^1$Departament de F\'{\i}sica Fonamental, Universitat de
 Barcelona, Diagonal, 647, 08028-Barcelona, Spain}

\address{$^2$Instituto dei Sistemi Complessi del CNR-Dipartimento di Fisica dell'Universit\`a
 di Roma ``La Sapienza'', P.le A. Moro 2, 00185 Roma, Italy}

\address{$^3$Department of Physics, 
 Norwegian University of Science and Technology (NTNU), NO-7491 Trondheim, Norway
 Present address: Institute for Traffic and Economics, Dresden University of Technology, D-01086 Dresden, Germany.}
 
\ead{josep.perello@ub.edu, miquel.montero@ub.edu, Luigi.Palatella@roma1.infn.it, ingves@phys.ntnu.no and jaume.masoliver@ub.edu}

\date{Received: \today / Revised version: \today}

\begin{abstract}
The electricity market is a very peculiar market due to the
 large variety of phenomena that can affect the spot price. However,
 this market still shows many typical features of other speculative
 (commodity) markets like, for instance, data clustering and mean
 reversion. We apply the diffusion entropy
 analysis~(DEA) to the Nordic spot electricity market (Nord Pool). We study 
 the waiting time statistics between consecutive
 spot price spikes and find it to show {\em anomalous} scaling
 characterized by a decaying power-law. The
 exponent observed in data follows a quite robust relationship with
 the one implied by the DEA analysis. We also in terms of the
 DEA revisit topics like clustering, mean-reversion and periodicities. We finally
 propose a GARCH inspired model but for the price itself. Models in the context of
 stochastic volatility processes appear under this scope to have a
 feasible description.
 \end{abstract}


\pacs{89.65.Gh, 05.45.Tp, 05.40.-a}


\vspace{2pc}

\noindent{\it Keywords}: Stochastic processes, Models of financial markets, Financial instruments and regulation

%\pacs{ {89.65.Gh}{Economics; econophysics, financial markets,
%business and management} \and {05.45.Tp}{Time series analysis}
%\and {05.40.-a}{Fluctuation phenomena, random processes, noise,
%and Brownian motion} } }
%
%\authorrunning{Perell\'o {\it et al.}}
\maketitle

\section{Introduction}

During the last years, there has been an increasing number of
contributions from the physics community to the study of economic
systems. Energy spot prices, that result from the deregulation of the
power sector, are no exception. Weron and Przybylowicz~\cite{weron3}
and Weron~\cite{weron1} deal with the Hurst exponent (or R/S)
analysis~\cite{RS}, the detrended fluctuation
analysis~(DFA)~\cite{DFA} and periodogram regression methods. These
techniques were used to verify and quantify a claim already stated by
financial mathematics: electricity spot prices are mean
reverting~\cite{schwartz,Book:Clewlow-2000}. This means that they
suffer a strong restoring force driving the price toward a certain
normal (``fundamental'') level. Using the language of physics, one
says that prices are {\em anti-persistent}, or equivalently, that the
price increments are negatively correlated. 

Recently, also the Average
Wavelet Coefficient~(AWC) method~\cite{AWC} has been applied to spot
prices~\cite{simonsen}. This method shows its potential in particular
when dealing with multi-scale time series. Due to its separation of
scale property, the presence of one scaling regime covering a given
time range is not hampered by the presence of another one. This is not
always the case for many other method, like the DFA, where one scaling
behavior can ``spill-over'' to the next one and even fully destroy the
scaling property of the latter one~\cite{simonsen}. In terms of power
markets, this is highly relevant since the statistical behavior of the
price on an intra-day scale is mainly determined by the consumption
patterns, and does not (on this scale) show the characteristic mean
reversion character that can be observed on the day-to-day scale, and
above~\cite{simonsen}. The lack of possibility in separating the
various scaling regimes, with the technique used, was the main
motivation for {\it e.g.} Refs.~\cite{weron3,weron1} analyzing mean
daily data, which show only one scaling region, instead of the
original hourly data. 

All the analyzing techniques mentioned above relay on the scaling of
some kind of fluctuation measure, say, the standard deviation or
variance, as function of the window size. Such approaches will only
measure the correct correlation scaling exponent, {\it i.e.} the
Hurst exponent, if the underlying time series is consistent with
Gaussian statistics~\cite{grigolinipre}. This is for instance the case
for the celebrated fractional Brownian motion~\cite{Ralf}. For correlated non-Gaussian
increments\footnote{It is here meant that the tails of the
 distribution are fatter than those of the Gaussian distribution.}
the associated exponent will partly receive contributions from the
correlations as well as the non-Gaussian character, making it
difficult (or impossible) to separate the two~\cite{grigolinipre}.
This latter situation is faced in {\it e.g.} ordinary L\'evy walks
and in its fractional equivalent~\cite{Ralf}. Recently, a method was
proposed that in a reliable way can determine and separate the
contribution to the scaling from both correlations and non-Gaussian
statistics. This method is based on the thermodynamics of the time
series and known as the {\em diffusion entropy analysis}
~(DEA)~\cite{giacomo-2,giacomo}.

In this paper we apply the DEA for the study of the electricity spot
prices from the Nordic Power exchange (Nord Pool). One of our main aims is to
investigate the statistics of the waiting times between consecutive
spot price spikes of deregulated electricity markets, and to uncover
if they show anomalous non-Poissonian statistics. The DEA is specially
suited for intermittent signals, {\it i.e.,} for time series where
bursts of activity are separated from periods of quiescent and regular
behavior. The technique has been designed to study the time
distribution of some markers (or events) defined along the time series
and thus discover whether these events satisfy the independence
condition~\cite{giacomo-2,giacomo}. Other objectives are the study
of the antipersitency behaviour in data via the DEA technique, 
the observation of existing periodicities in data in many different ways 
and finally the proposal of a GARCH model showing properties similar to those of real data.

This paper is organized as follows. We start in Sec.~\ref{NordPool} by
briefly discussing the spot market and the data set to be analyzed in
this work. Then we briefly review the DEA technique~(Sec.~\ref{DEA}).
Section~\ref{spikes} studies the statistics of the most important
price movements and infer some properties on the inter-event peak
probability. Section~\ref{mean} presents the results obtained by the
DEA technique on tick-by-tick spot data and without filtering the most
relevant price changes. Furthermore, we propose in Sec.~\ref{garch} a
GARCH model~\cite{escribano,engle,bollerslev,engle-patton,duffie} for
the spot electricity price and try to obtain consistent entropies for
the tick-by-tick data and the spike filtering. Conclusions are left
for Sec.~\ref{conclusions}.

\section{Nord Pool --- the Nordic power exchange\label{NordPool}}

The Nordic commodity market for electricity is known as Nord
Pool~\cite{NordPool}. It was established in 1992 as a consequence of
the Norwegian energy act of 1991 that formally paved the way for the
deregulation of the electricity sector of that country. At this time
it was a Norwegian market, but in 1996 and 1998 Sweden and Finland
joined, respectively. With the dawn of the new millennium (2000),
Denmark decided to become member as well.

Nord Pool was the world's first international power exchange. In this
market, participants from outside the Nordic region are allowed to
participate on equal terms with ``local'' participants. To
participate in the spot market it is required that the participants
must have an access to a grid connection enabling them to deliver or
take out power from the main grid. For this reason, the spot market is
often also called the physical market. As for today, the physical
market has a few hundreds of participants. More than one third of the
total power consumption in the Nordic region is traded in this market,
and the fraction has steadily been increasing since the inception of
the exchange in the early 1990s. In addition to the physical market,
there is also a financial market. Here, power derivatives, like
forwards, futures, and options are being traded. This market
presently has about four hundred participants. For each of these
two markets, about ten nationalities are being represented among the
market participants.

Nord Pool is an auction based market place where one trades power
contracts for physical delivery within the next day. It is known as a
{\em spot market}. However, in a strict sense this notion is not
precise since formally it is a day-ahead ($24$-hours) forward market.
What is traded are one-hour-long physical power contacts, and the
minimum contract size is $0.1$MWh. By noon (12:00 hours) every day,
the market participants submit their (bid and ask) offers (including
prices and volumes) to the market administrator (Nord Pool). The
offers are submitted for each of the individual $24$ hours of the next
day starting at 1:00 hours. After the submission deadline (for the
next day), Nord Pool proceeds by preparing (for ever hour) cumulative
volume distributions (purchase and sale curves) {\it vs.} price ($p$)
for both bid~($V_B(p)$) and ask~($V_A(p)$) offers. Since there in the
electricity market must be a balance between production and
consumption, the so-called {\em system (spot) price}, $P(t)$, for that
particular hour ($t$) is determined as the price where $V_A(P)=V_B
(P)$. This is called the {\em market cross}, or {\em equilibrium
 point}. Trading based on this method is called equilibrium trading,
auction trading or simultaneous price setting. If the data do not
define an equilibrium point, no transactions will take place for that
hour, and no system spot price will therefore be determined. So far,
to our knowledge, this has never happened at Nord Pool.

%------------------------------------------------------------
\begin{figure}[t]
 \begin{center}
 \includegraphics*[width=0.92\columnwidth]{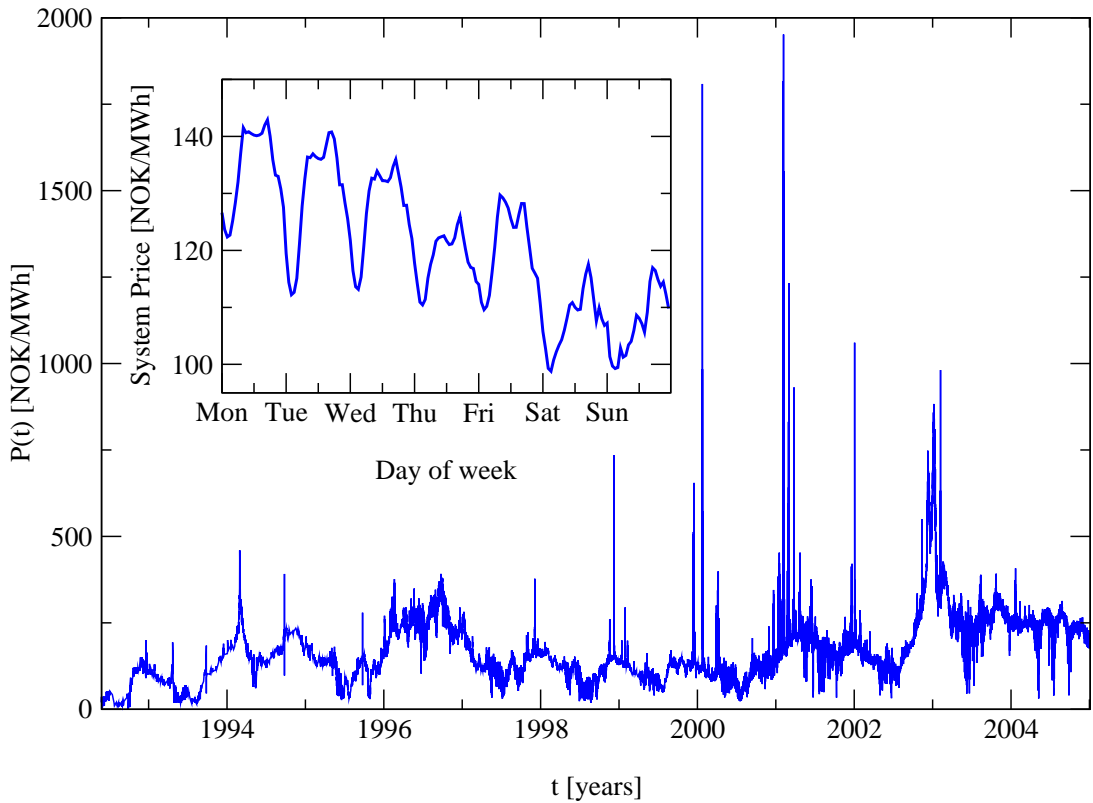}
 \caption{\label{fig1} The Nord Pool system spot price from May,
 1992 till the end of 2004. Several (seasonal, weekly, and daily)
 periodicities can be observed in this data set below the spiky
 randomness. The inset shows the price variations over a
 (randomly chosen) weekly period.}
 \end{center}
\end{figure}
%------------------------------------------------------------

After having determined the system price, $P(t)$, for a given next-day
hour, Nord Pool looks for potential {\em bottlenecks}
(grid congestions) in the power transmission grid that might result
from this system price. If no {\em bottlenecks} are found, the system
price will represent the spot price for the whole Nord Pool area for
that given hour. On the other hand, if potential grid congestions may
result from the bidding, so-called {\em area (spot) prices}, that are
different from the system price, will have to be created. The idea
behind the introduction of area prices is to adjust electricity prices
within a geographical area in order to favor {\em local} trading to
such a degree that the limited capacity of the transmission grid is
not exceeded. How the area prices are being determined within Nord
Pool differs between, say, Sweden and Norway, and we will not discuss
it further here (see {\it e.g.} Ref.~\cite{NordPool} for details).

In this work, we will analyze the {\em hourly} Nord Pool system spot
prices for the period from (Monday) May 4th, 1992 till the end of
Friday December 31st, 2004; in total 110,987 data
points~(Fig.~\ref{fig1}). In Fig.~\ref{fig1a} we present the
corresponding normalized hourly returns (to be defined in
Eq.~(\ref{log}) below). The reader should notice that high levels of
return are possible, and typical, for electricity markets. From
Fig.~\ref{fig1} one should also observe that the price process shows
several periodicities. Those are mainly attributed to consumption
patterns and are daily, weekly and seasonal in character. They have
been reported and studied previously by several research
groups~\cite{simonsen,escribano,Volatility,pilipovic,schwartz1}
Superimposed onto this deterministic periodic trend, is a random
component showing strong variability with pronounced spikes and data
clustering. The DEA technique allows us to deal with this randomness
by somewhat ignoring the periodic signal.

%------------------------------------------------------------
\begin{figure}[t]
 \begin{center}
 \includegraphics*[width=0.92\columnwidth]{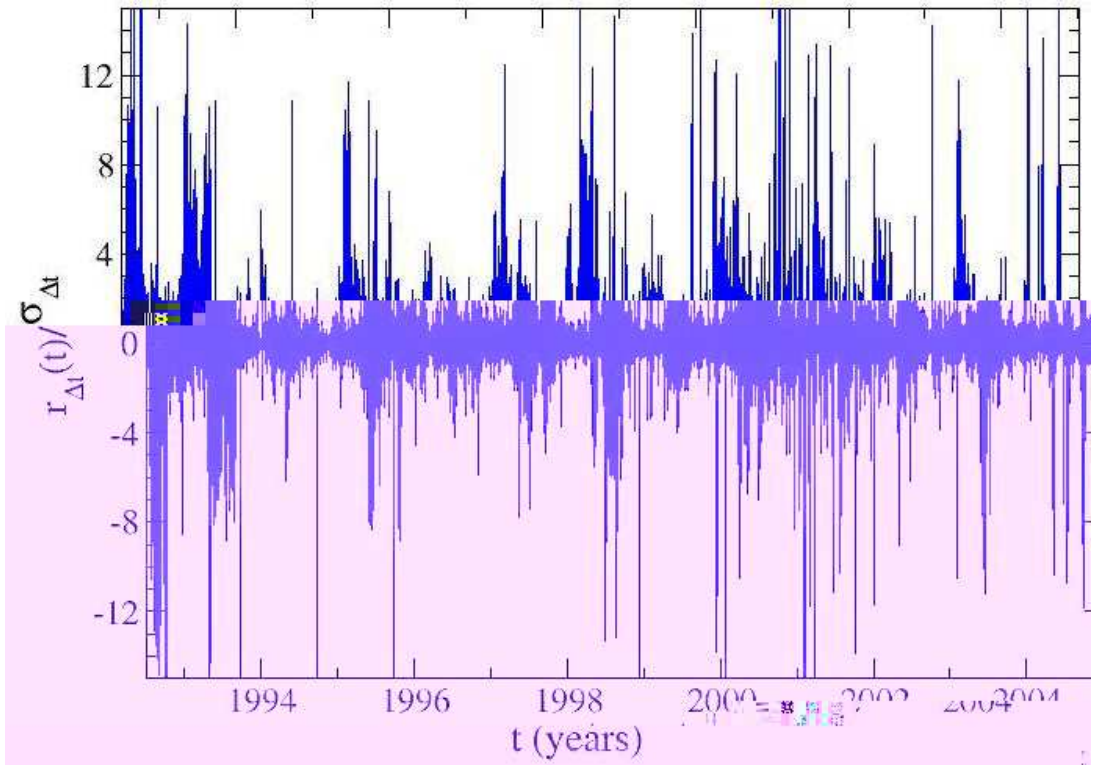}
 \caption{\label{fig1a} The normalized Nord Pool hourly ($\Delta t =
 1$~hour) (logarithmic) returns, $r_{\Delta t}(t)/\sigma_{\Delta
 t}$ ({\it cf.} Eq.~(\ref{log})), of the data depicted in
 Fig.~\protect\ref{fig1}. The normalization is done according
 to the average hourly volatility $\sigma_{\Delta t}$. Notice
 that normalized return values higher than $10$ is rather
 frequent and typical for this (and many other) electricity
 markets.}
 \end{center}
\end{figure}
%------------------------------------------------------------

\section{The Diffusion Entropy Analysis\label{DEA}}

The DEA technique is a statistical method for measuring scaling
exponents of time series by utilizing their thermodynamical
properties~\cite{giacomo-2,giacomo}. This is achieved by {\it (i)}
converting the time series into some kind of probability density
function (pdf), $p(x,t)$, where the variable $x$ is related to the
fluctuating time series and $t$ denotes the time (or time interval),
and {\it (ii)} therefrom calculating the related Shannon (information)
entropy $S(t)$ (to be defined below), from which the scaling
properties (if any) of the pdf can be deduced ({\it cf.}
Eqs.~(\ref{Scaling}) and (\ref{delta}) below).

Various pdf's of the form $p(x,t)$, obtained from the underlying time
series, can be used together with the DEA technique. For instance,
$p(x,t)$ can be the probability of being at position $x$ at
(diffusion) time $t$ {\em given} that one was at $x=0$ at time $t=0$,
or $x$ may be a ``marker'' related to the number of events (like
values of the time-series being above a given threshold {\it etc.})
occurring within a time interval of length $t$. Having said that, it
should be mentioned that generally $x$ should be a zero-mean variable
in order to avoid artifacts in the application of the
DEA~\cite{giacomo-2,giacomo}. We will in the preceding section
describe in more details how one may construct a suitable pdf to be
used in the DEA.

Under the assumption that the time series is stationary and scale
invariant, one has that the pdf can be written as
\begin{equation}
\label{Scaling}
p(x,t)=\frac{1}{t^\delta}\,F\left(\frac{x}{t^\delta} \right),
\end{equation}
where $\delta$ denotes the so-called {\em scaling exponent} (that one
wants to determine), and $F$ is some positive and integrable function
depending on the specificities of the pdf. Then the Shannon entropy,
that in general is defined as
\begin{equation}
\label{Shannon}
S(t) = - \int_{-\infty}^{\infty} \mathrm{d}x \; p(x,t)\, \ln[p(x,t)],
\end{equation}
will take on the form
\begin{equation}
\label{delta}
S(t) = A+ \delta \ln t,
\end{equation} 
where $A$ is given by an expression similar to Eq.~(\ref{Shannon}),
but with $F(y)$ substituted for $p(x,t)$. This last transition is
easily demonstrated by substituting Eq.~(\ref{Scaling}) into
Eq.~(\ref{Shannon}) and making a change of variable to $y=x/t^\delta$.

The important lesson that should be taken from Eq.~(\ref{delta}) is
that the scaling exponent, $\delta$, can be obtained as the slope of
the entropy {\it vs.} $\ln t$ curve. This is a reliable method
\cite{grigolinipre,giacomo-2,giacomo} for measuring the scaling
exponent of the probability density~(\ref{Scaling}).

\section{Spikes and data clustering\label{spikes}}

One of the most pronounced features of the electricity spot price
process of Fig.~\ref{fig1} is its spiky nature. Within one, or only a
few hours, the spot price can increased manifold. One of the most
extreme example of such a situation occurred on February 5, 2001, when
the spot price reached an all-time-high of $1951.76$ NOK/MWh. Three
hours earlier, the spot price had been at about $200$ NOK/MWh, a more
``normal'' level for that time of year.

%------------------------------------------------------------
\begin{figure}
 \begin{center}
 \includegraphics*[width=0.92\columnwidth]{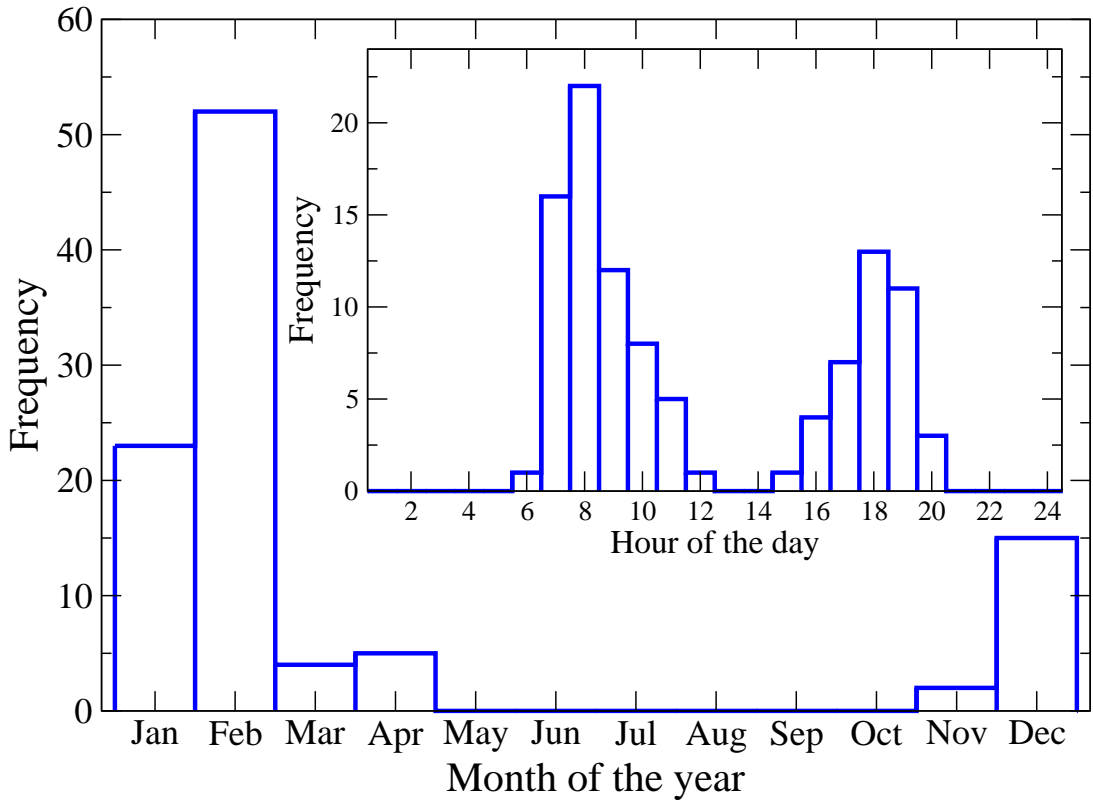}
 \caption{\label{fig1c} The frequency distributions of the $100$
 largest hourly return events {\it vs.} month of year and hour of
 the day (inset) based on the data depicted in
 Fig.~\protect\ref{fig1a}. This result clearly indicates that the
 dramatic price changes are not evenly distributed in time, but
 instead occur during periods of high consumption.}
 \end{center}
\end{figure}
%------------------------------------------------------------

The dramatic price variations take place during periods of high
consumption, which for the Nordic area means during the winter season
as illustrated by Fig.~\ref{fig1c}. One way to explain the appearance
of these dramatic price changes is by the so-called {\em stacking or 
inventory models}~\cite{weron}: Extra demand for electric power is
normally filled by the cheapest available energy source. During low
consumption periods, the daily consumption profiles, say, will not
influence the price level in any dramatic way, since there is plenty
of base energy generation capacity available. However, when the
consumption already is high, only a minor increase in the demand can
have dramatic consequences for the electricity prices. The higher the
demand, the more costly will it typically be to produce an extra unit
of energy, and the more sensitivity will the price of electric power
be to minor changes in the consumption. If one during such
high-consumption periods experiences loss of less costly generating
capacity, {\it e.g.} due to technical problems, extreme situations like
the one reported above can be experienced. This sensitivity of the
spot price to the level of consumption implies seasonal volatility
clustering with typical low and high volatility periods during the
summer and winter seasons, respectively~\cite{Volatility}.

The fact that the extreme price spikes are more likely to occur during
the winter season than during the remaining parts the year, has the
immediate consequence that time intervals between (consecutive) spikes
are not expected to constitute a Poisson process. Hence, the time
intervals are expected to scale anomalously according to
Feller~\cite{Feller}, and hence the scaling exponent will be different
from $1/2$. This waiting time statistic (between spikes) has
previously been studied in {\it e.g.} Ref.~\cite{grigolinipre} for
solar flares or earthquakes. It is worth mentioning that although the 
presence of seasonalities in spike statistics is quite evident looking at
Fig.~\ref{fig1c}, this does not mean that the spike events follow a somewhat 
deterministic rule. As we will show later, time between spikes $\tau$ 
appears to be described by a power law probability distribution.
The periodicities affect the statistics of 
very particular values of $\tau$ but not the whole probability density function.

It has in the past been demonstrated~\cite{giacomo} that if the
``spike'' events are {\em independent} and the time interval between
them distributed (at least asymptotically) according to a power law of
the form ($\beta>0$)
\begin{equation}
\phi(\tau)\sim 1/\tau^\beta,
\label{waiting}
\end{equation}
then an analytic relationship between the $\beta$ and the scaling exponent $\delta$ do exist. For
instance, if one constructs an auxiliary process by introducing steps,
say, of magnitude $1$, whenever the originally set has absolute
returns larger than a given threshold (called a marker below), the
relationship reads~\cite{giacomo}
\begin{equation}
 \delta_{\rm spikes}= 
 \cases{ \beta-1, & if \ $1<\beta\leq 2$;\cr
 1/(\beta-1), & if \ $2<\beta\leq 3$;\cr
 1/2, & if \ $\beta>3$.}
\label{giacomoeq}
\end{equation}

Under this scope, the present section investigates the predictive
power of the DEA analysis applied to the spike statistics of the
Nordic electricity market. From the hourly system spot prices data,
$P(t)$, one defines the logarithmic returns\footnote{The reader should
 notice that due to the large price variations (relative to the price
 level itself) that can occur for electricity spot prices, {\em
 relative} and {\em logarithmic} returns are not necessarily even
 close to being similar. This is in sharp contrast to, say, stock
 markets where the two types of returns are approximately equal in
 most cases.} over the time horizon, $\Delta t$, as
\begin{equation}
 \label{log}
 r_{\Delta t}(t) %\equiv \ln P(t)-\ln P(t-\Delta t)
 \equiv \ln\left( \frac{P(t)}{P(t-\Delta t)} \right),
\end{equation}
and in the following it will be assumed that $\Delta t =1$ hour. We
intend to investigate the properties and time distribution of those
returns being larger than a certain lower threshold $r_0$. To this
end, we define a time-dependent marker position variable by
\begin{equation}
 \label{eq:marker}
 \xi(t)=\cases{1, &if \ $|r_{\Delta t}(t)|>r_0$,\cr
 0, &otherwise.}
\end{equation}
One now wants to estimate (count) the number of markers,
$\tilde{y}_t(T)$, over a time interval $T$ ending at time $t$.
With Eq.~(\ref{eq:marker}), this can readily be done by 
\begin{equation}\label{movingcounting}
 \tilde{y}_t(T)=\sum \limits_{i=t-T}^{t} \xi(i),
 \label{y(t)1}
\end{equation}
and by subtracting the arithmetic average of this process one gets
\begin{equation}
 y_t(T)= \tilde{y}_t(T)-\left< \tilde{y}_t(T) \right>_t.
 \label{y(t)}
\end{equation}
If we vary the value of $t$ along the interval $[T,N\Delta t]$, where
$N$ is the total number of data points of the analyzed sequence. From
the knowledge of $y_t(T)$, one can now readily calculate the
probability density function $p(y,T)$ for the moving counter.

\begin{figure}
 \begin{center}
 \includegraphics[width=0.92\columnwidth]{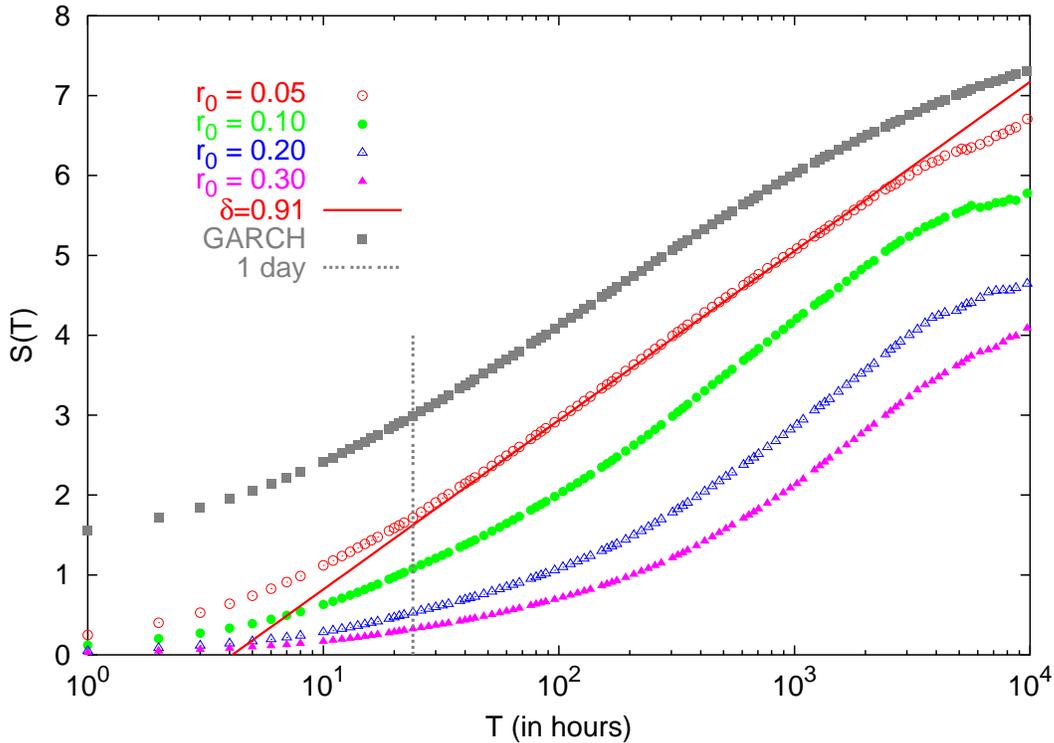}
 \caption{\label{fig2} The results of the DEA of the Nord Pool spot
 price data using as marker $|r_{\Delta t}(t)|>r_0$ (with $\Delta
 t=1$~hour) for different thresholds $r_0$. The value of
 $\delta_{\rm spikes}$ is practically constant and, therefore,
 periodicities do not affect this kind of data analysis. For the
 GARCH curve, we take $r_0=0.2$ and we shift the results to see
 clearly. We have also added a vertical line to distinguish the intraday regime.
 The entropy there is highly affected by the consumption patterns of human activity.}
 \end{center}
\end{figure}

At this point we apply the DEA technique~\cite{giacomo-2,giacomo}.
For each time-window of size $T$, one calculates the probability
density function of $y_t$, that is: we get $p(y,T)$ for several $T$'s.
The related (Shannon) entropy, $S(T)$, is then defined in accordance
with Eq.~(\ref{Shannon}). In Fig.~\ref{fig2} we present the behavior
of the Nord Pool entropy (using the marker defined in
Eq.~(\ref{eq:marker})) for different values of threshold $r_0$. It is
observed that the entropy, $S(T)$, scales in accordance with
Eq.~(\ref{delta}), and that the scaling exponent is found to be
around $\delta_{\rm spikes} =0.91$ for all thresholds (solid line in Fig.~\ref{fig2}). 
See Table~\ref{tab1} for further details on the fitting procedure.

It is interesting to notice that
a similar value (within the error bars) for the scaling exponent has
previously been reported for the activity of the US Dollar--Deutsche Mark
futures market (using tick-by-tick data)~\cite{palatella} as well as
for daily index data from the Dow Jones Industrial
Average~\cite{palatella1}. Hence, the Nordic power market has similar
entropy growth to these two markets. Furthermore, all these data sets
have comparable clustering.

%------------------------------------------------------------
%------------------------------------------------------------

%------------------------------------------------------------
\begin{table}
 \caption{Regressions for the diffusion entropy results 
 given in Fig.~\ref{fig2} assuming an entropy of the form $S(t)=A+\delta_{\rm spikes}\ln t$. 
 Last column gives the time domain over which the regression is made.}
 %\begin{center}
 \begin{indented}
 \item[]
 \begin{tabular}{llcc}
 \hline
 \hline
 $r_0$ & $\delta_{\rm spikes}$ & {\footnotesize data points} & {\footnotesize time range in hours}\\
 \hline
 0.05 & $0.913 \pm 0.002$ & 57 & $40-2,000$ \\
 0.10 & $0.911\pm 0.008$ & 47 & $100-2,000$ \\
 0.20 & $0.918 \pm 0.008$ & 33 & $200-2,000$ \\
 0.30 & $0.91\pm 0.01$ & 20 & $500-2,000$\\
 \hline
 \hline
 \end{tabular}
 %\end{center}
 \end{indented}
 \label{tab1}
\end{table}
%------------------------------------------------------------

As was alluded to in the introduction to this section, the DEA
technique aims at measuring the inhomogeneity of the distribution of
the number of events (as defined by a marker) over a fixed period of
time $T$, and to obtain the related scaling exponent $\delta$. 
An alternative (indirect) approach to obtaining the
scaling exponent is to measure the tail exponent $\beta$ of the
waiting time distribution $\phi(\tau)$, and apply relation~(\ref{giacomoeq}) 
in order to obtaining the scaling exponent (under the stated assumptions).

\begin{figure}[t]
 \begin{center}
 \includegraphics[width=0.92\columnwidth]{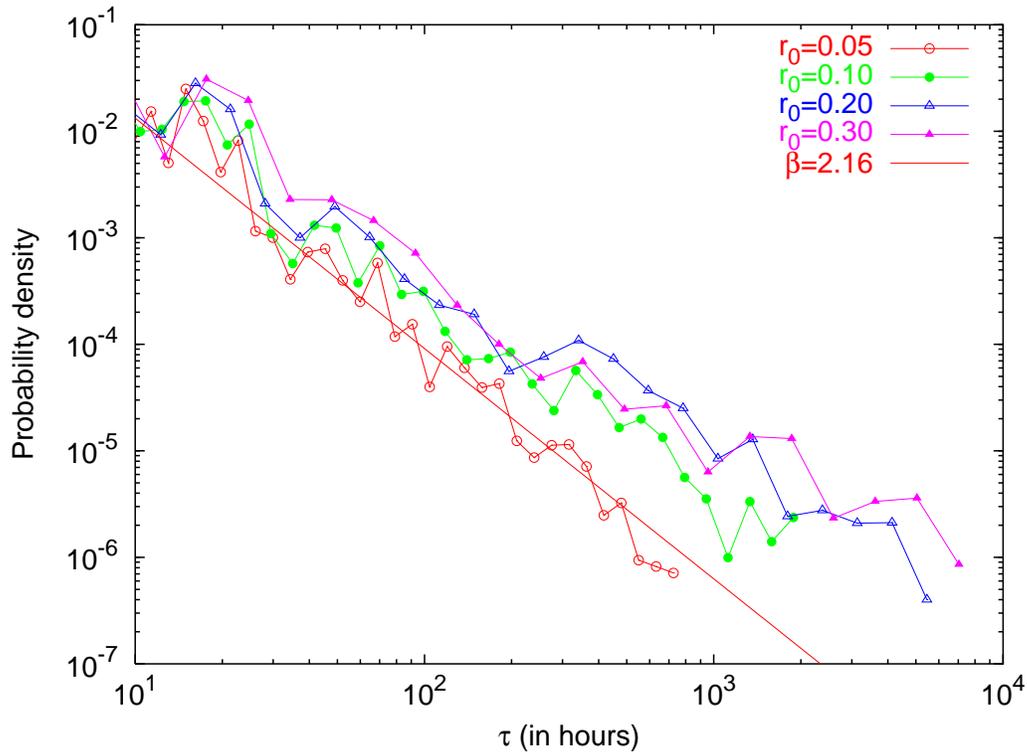}
 \caption{\label{fig2a} The waiting time pdf, $\phi(\tau)$, {\it
 vs}. the time interval between spikes, $\tau$, defined
 according to $|r_{\Delta t}(t)|>r_0$ (with $\Delta t=1$~hour)
 for different values of the threshold $r_0$ as indictaed in the
 legend. The data collapse for the different curves towards a
 decaying power law, $\phi(\tau)\sim 1/\tau^\beta$, is reasonably good. 
 The individually measured waiting time exponents $\beta$ are given in
 Table\protect~\ref{tab}. The solid line corresponds to a
 power-law fit with $\beta=2.16\pm 0.07$ for the threshold $r_0=0.05$.}
 \end{center}
\end{figure}

Figure~\ref{fig2a} depicts the pdf of waiting times between two
consecutive spikes, $\phi(\tau)$, for different choices of the
threshold $r_0$. It is mostly consistent with a decaying
power-law of the form (\ref{giacomoeq}).
There seems to be a slight dependence of the waiting time exponent
$\beta$ on the choice of threshold, particularly for the largest
waiting times $\tau$. One in Table~\ref{tab} therefore presents the
measured value of $\beta$ for the thresholds used. The corresponding 
scaling exponent obtained according to
Eq.~(\ref{giacomoeq}), is given in the last column of Table~\ref{tab}.
One notices that the two methods --- the direct {\it via} the diffusion 
entropy, and the indirect determination from the spike waiting time distribution 
--- are consistent within the
error-bars.

%------------------------------------------------------------
%------------------------------------------------------------

%------------------------------------------------------------
\begin{table}
 \caption{Regressions for the tail of the waiting time pdf 
 $\phi(\tau)\sim1/\tau^\beta$ between spikes larger than a certain threshold $r_0$. The fit 
 provides the estimated exponent $\beta$ (see Eq.~(\ref{waiting})) 
 over the data plotted in Fig.~\ref{fig2a}. Last column corresponds 
 to the scaling exponent $\delta$ predicted from the measured
 values of $\beta$ and assuming Eq.~(\ref{giacomoeq}) to be right.}
 %\begin{center}
 \begin{indented}
 \item[]
 \begin{tabular}{llcl}
 \hline
 \hline
 $r_0$ & $\beta$ & {\footnotesize data points} & {\footnotesize Predicted $\delta_{\rm spikes}$}\\
 \hline
 0.05 & $2.16\pm0.07$ & 39 & $0.86\pm 0.05$ \\
 0.10 & $2.10\pm0.30$ & 11 & $0.85\pm 0.17$ \\
 0.20 & $1.84\pm 0.15$ & 11 & $0.84\pm 0.15$ \\
 0.30 & $1.79\pm 0.12$& 11 & $0.79\pm 0.12$\\
 \hline
 \hline
 \end{tabular}
 %\end{center}
 \end{indented}
 \label{tab}
\end{table}
%------------------------------------------------------------

\section{Mean reversion and consumption patterns\label{mean}}

There is in the literature wide consensus on the fact that electricity
prices are stationary in the sense of suffering a reverting force
driving price towards a normal
level~\cite{weron3,weron1,schwartz,Book:Clewlow-2000,simonsen,pilipovic,schwartz1,weron,weron4}.
This level is most often time dependent due to marked consumption
pattern in electricity prices caused by human social activity, and
climatic factors like temperatures (see {\it e.,g.} Fig.~\ref{fig1c}).
We will now for consistency address the mean reverting character of
the Nordic spot electricity prices, but now using the DEA technique.
Previously, using the AWC techniques, the so called Hurst exponent,
$H$, characterizing this property, has been measured to be about
$0.4$~\cite{simonsen}, a similar value also found for the Californian
electricity market~\cite{weron3}.

For preparing for the use of the DEA for the measurement of the Hurst
exponent, we define simply the ``marker'' being equal to the hourly
returns, that is 
\begin{equation}
 \xi(t) = \ln(P(t)/P(t-\Delta t))= r_{\Delta t}(t).
 \label{marker2}
\end{equation}
The corresponding moving counting then, in analogy with
Eqs.~(\ref{movingcounting}) and (\ref{y(t)}), reads
\begin{equation}
 y_t(T)=\sum \limits_{i=t-T}^{t} r_{\Delta t}(t) = \ln P(t)-\ln P(t-T)=r_T(t),
 \label{marker21}
\end{equation}
where one in the last transition has used the definition of the
return (\ref{log}).

\begin{figure}[t]
 \begin{center}
 \includegraphics[width=0.92\columnwidth]{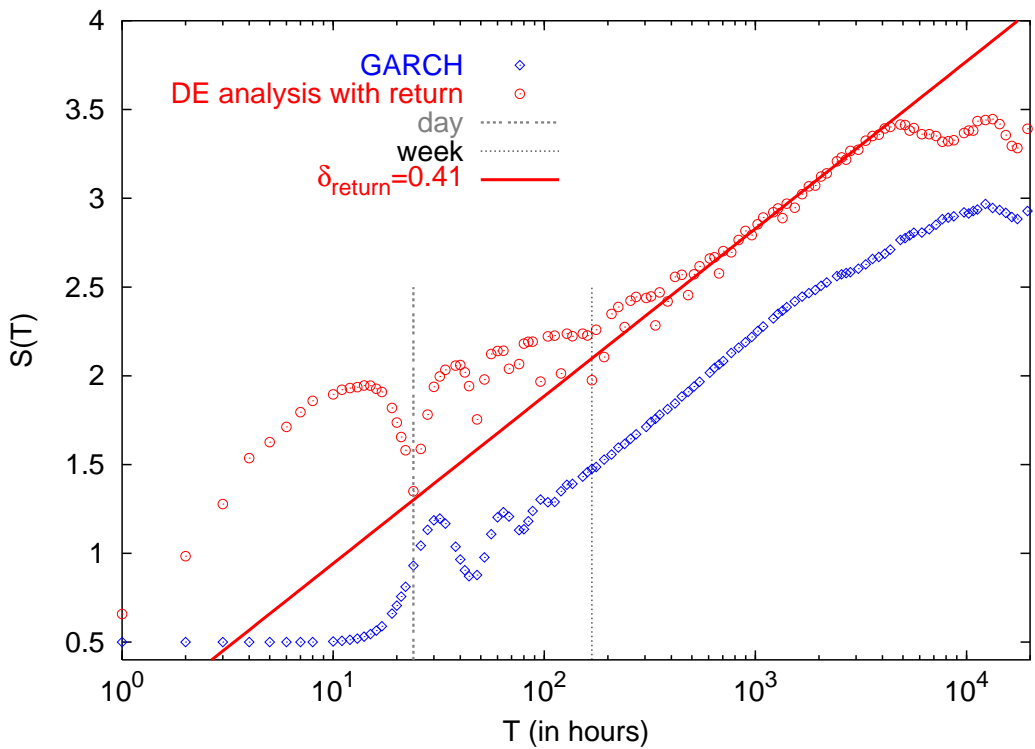}
 \caption{\label{fig3} The result of DEA using $\xi(t)=r_{\Delta t}(t)$
 as marker. The vertical lines signal the time of the main
 periodicities. We perform the analysis with the empirical data
 and a simulation of a GARCH model~(\ref{eq:garch}). The two time series have the same growth 
 with time although for short time lags entropy differs due to the presence of periodicities in the original electricity spot prices.}
 \end{center}
\end{figure}

Figure~\ref{fig3} presents the (Shannon) entropy~(\ref{Shannon}) of
the moving counter $\xi(t) = r_{\Delta t}(t)$. It is observed that
the stationary state is reached after about $5,000$ hours which 
corresponds to almost seven months. The signal will diffuse anomalously 
until it feels that the motion of $r_{\Delta t}(t)$ is constrained by the reverting force.
After this threshold, the entropy $S(T)$ remains constant and the
asymptotic scaling will be $\delta=0$ (cf. Eq.~(\ref{delta})) and the
signal loses its memory.

According to Fig.~\ref{fig3} there also exists a transient state for
shorter times than $T=5,000$ hours. In this regime, the DEA detects the 
presence of daily and weekly
periodicities but it seems to be insensitive to seasonalities. The
analysis, excluding the periodicities, also allows a fit like the one
given by Eq.~(\ref{delta}), that is
\begin{equation}
 S_{\rm fit}=A+\delta_{\rm return}\ln T,
 \label{sreturn}
\end{equation}
where in this case $\delta_{\rm return}$ equals the Hurst exponent $H$
use to the hourly returns being used as markers. From Fig.~\ref{fig3}
it is observed that the DEA calculated entropy can be well fitted
in time by the Hurst exponent
\begin{equation}
H_{\rm DEA}=\delta_{\rm return}= 0.411\pm 0.002,
\end{equation}
for the range between 100 and 6,000 hours. This result fully coincides 
with the Hurst exponent measured previously with the AWC method~\cite{simonsen}. 
If the Hurst exponent $H=\delta_{\rm return}$ was equal to $1/2$ we would talk about a
(ordinary) diffusion process with a standard deviation growing like
$\sqrt{T}$. Therefore, finding a value smaller than $1/2$ for the
Hurst exponent $H=\delta_{\rm return}$ means that the process is
antipersistent or anticorrelated~\cite{Feller}.

\begin{figure}[t]
 \begin{center}
 \includegraphics[width=0.92\columnwidth]{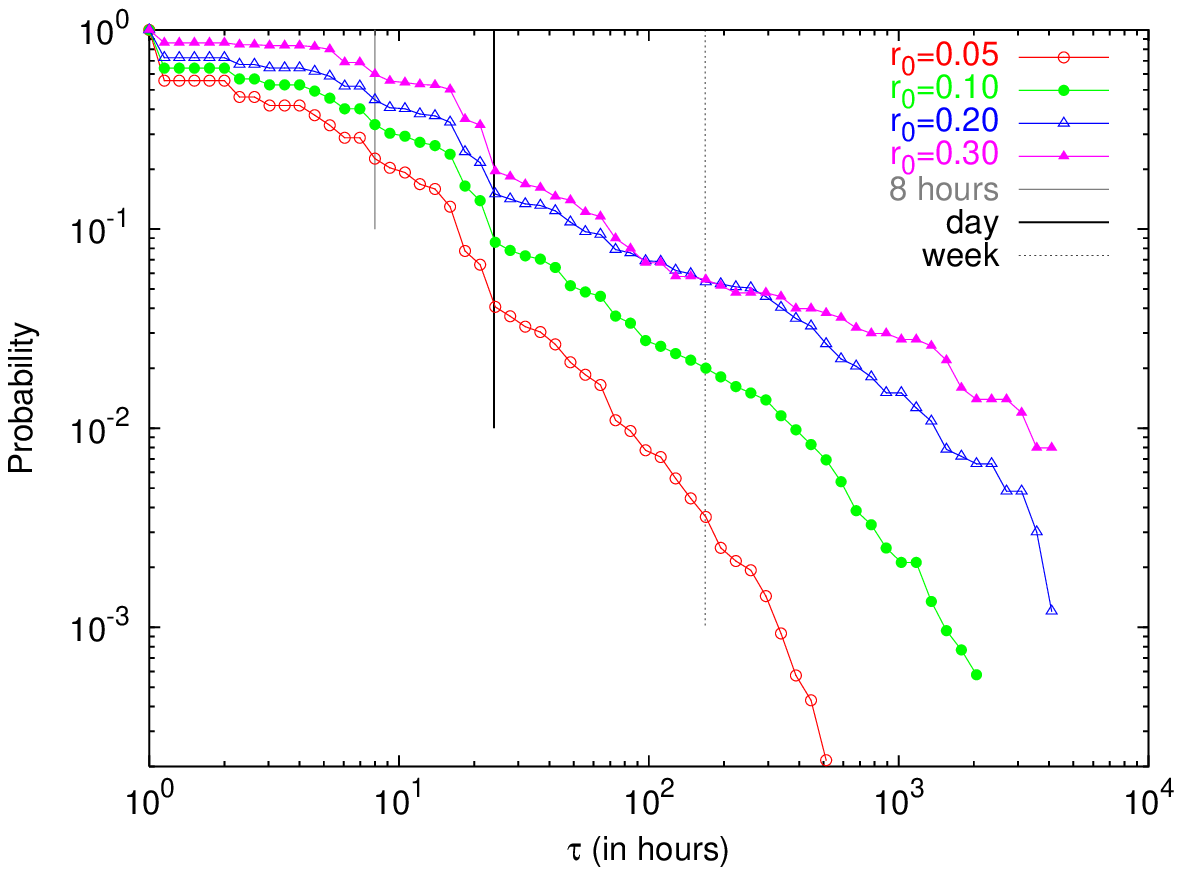}
 \caption{\label{fig3b} The
 complementary cumulative distribution of inter-spike waiting
 times, $\Phi(\tau)=1-\int_\infty^\tau d\tau' \phi(\tau')$. 
 One observes that the decay of these distributions are different in several 
 regions. This can be attributed to the presence of periodicities in the system spot
 prices. The slope changes are important in 8 hours and day (24 hours) time 
 lags but are not unobservable in a week time lag.}
\end{center}
\end{figure}

Going further into the issue of periodicities, we now again explore
the distribution of spikes distance $\tau$ for different thresholds.
We display in Fig.~\ref{fig3b} the complementary cumulative
probability of this random variable. Compared to probability density
plotted at Fig.~\ref{fig2a} we note that the complementary cumulative
distribution function of the spikes distance is greatly distorted by
the presence of periodicities in the underlying data which breaks a
possible power law collapse for several decades. This is mainly due to
the gap in eight hours and daily periodicities while weekly effects
does not affect very much the cumulative curve. These distortions are
consistent with the periodicities detected by the DEA technique. It is
again difficult to see the seasonal effects but clearly detect the
rest of periodicities.

\section{A GARCH type model\label{garch}}

We have seen that the spike statistics for the electricity prices is
very similar to that of the Dow Jones index and also to the peak
statistics of the US Dollar--Deutsche Mark futures
market~\cite{palatella,palatella1}. They both have the same data
clustering. The analogy leads us to search for good candidates for
price models for electricity in a similar way to as was done
successfully in Refs.~\cite{palatella,palatella1}. Volatility clustering and mean reversion 
are among the essential characteristics of the volatility and of the
market activity. Henceforth, our survey focuses on existing
volatility models.

A possible new attempt is to propose a GARCH model~\cite{engle-patton}
for the spot price. We suggest the following model for the
system spot price $P(t)$:
\begin{equation}
 \label{eq:garch}
 P(t) = k + \chi P(t-\Delta t) \eta^2 + \nu P(t-\Delta t),
\end{equation}
where $\eta$ denotes uncorrelated gaussian random noise of zero mean
and unit variance. Using this GARCH theoretical model, with
parameters $\chi=4\times 10^{-4}$, $\nu=0.9994$ and $k=0.00184$ NOK/MWh, one
can simulate long time series that results in an $S(T)$ dependence
that resembles that of the real Nord Pool market (Fig.~\ref{fig2}). To obtain 
the results of Fig.~\ref{fig2} the total
length of the simulated and real data were the same, and that
the markers $|r_{\Delta t}(t)|>r_0$ were used. 
This model can generate $\delta_{\rm spikes}
\simeq 0.9$ thus giving similar entropy profile to the electricity spot price data.

In addition, we also performed the DEA for the logarithmic return time
series, generated by Eq.~(\ref{eq:garch}) using the parameters given
above, and markers defined in Eq.~(\ref{marker2}). 
Figure.~\ref{fig3} shows that this GARCH model produces results for the diffusion entropy that are
consistent with real data. Specifically, the transient period
resulting from the model time series has the same exponent and the
stationary state is approached simultaneously with the real electricity
return time series. Moreover, if Eq.~(\ref{eq:garch}) is rewritten in
terms of the price difference $\Delta P(t)=P(t)-P(t-\Delta t)$
\begin{equation}
 \Delta P(t) = -(1-\nu) \left[P(t-\Delta t) - \frac{k}{1-\nu}\right]
 + \chi P(t-\Delta t) \eta^2,
 \label{garch1}
\end{equation}
we can observe that the reverting force has a characteristic time
scale $\hat{\tau}=1/(1-\nu)$. It is directly related to $\nu$ and brings
us the value $\hat{\tau}=1,667$ hours (more than 2 months) which is of the
same order of the time required for real data to arrive at the
stationary state. The ratio $k/(1-\nu)$ represents the normal level that prices are driven toward. 
In case, one would like to include the consumption patterns
in the model one should replace $k$ by a periodic signal. Finally, the
$\chi$ value sets the magnitude of the fluctuations. Having a big $\chi$
means having wild fluctuations of price. 

\section{Conclusions\label{conclusions}}

By applying the DEA technique we have shown that the entropy of the
Nordic electricity spot prices grows with the size of the time window in a similar
way to the volatility and the market activity as different as the Dow-Jones and the 
US Dollar--Deutsche mark futures. The DEA scaling parameter has the similar value
$\delta_{\rm spikes}\sim 0.9$ and they have a comparable
clustering.

We have also shown that the distribution of waiting times of the
electricity price scales as $\phi(\tau)\sim 1/\tau^\beta$
given by Eq.~(\ref{waiting}) and that the corresponding 
power-law exponent $\beta$ is related to the DEA
parameter $\delta_{\rm spikes}=0.91\pm 0.01$ as anticipated by the DEA -- see Eq. (\ref{giacomoeq}) and Fig.
\ref{fig2a}. The relationship derived in Ref.~\cite{giacomo} is obtained under the hypothesis that spikes are uncorrelated. This hypothesis might hold true or not but in any case we have found that the relationship between $\beta$ and $\delta_{\rm spikes}$ is still applicable in the Nord Pool case. 

We have also obtained a Hurst exponent ($\delta_{\rm return}=0.412\pm 0.002$) 
with the DEA technique which is coincident with the one derived by previous studies 
and using different techniques. The resulting quantity implies an antipersistent 
behaviour of electricity prices. Moreover, we have been able to detect, for hourly
prices, a characteristic time scale around $5,000$ hours in which
the system reaches the stationary state.

Due to the several similarities in statistical properties with
financial markets, a GARCH model was proposed and investigated for the
spot electricity prices. In particular, it was shown that the
diffusion entropy of such a theoretical GARCH model followed
surprisingly closely the entropy that was obtained for the Nord Pool
system spot prices. 

We finally mention that although the DEA seems to detect daily and/or
weekly periodicities in the analyzed data, a more thorough
investigation is required to settle this issue. Efforts are currently
undertaken to address this topic.

\ack
%\section*{Acknowledgements}
%\begin{acknowledgement}
 JP, MM and JM want to acknowledge the support received by
 Direcci\'on General de Investigaci\'on under contract No.
 BFM2003-04574. The authors are grateful to SKM Market Predictor
 for providing the data analyzed in this paper. 
%\end{acknowledgement}

\end{document}